\documentclass[10pt,twocolumn,letterpaper]{article}

\usepackage{cvpr}
\usepackage{times}
\usepackage{epsfig}
\usepackage{graphicx}
\usepackage{amsmath}
\usepackage{amssymb}

\usepackage{amsfonts} 
\usepackage{cases}
\usepackage{overpic}
\usepackage{caption}
\usepackage{subcaption}
\usepackage{multirow}
\usepackage{tabularx}
\usepackage{enumitem}
\usepackage{flushend}

\def\bx{{\bf x}}
\def\by{{\bf y}}
\def\bz{{\bf z}}

\def\0{{\bf 0}}
\def\1{{\bf 1}}

\def\bC{{\bf C}}

\def\bH{{\bf H}}

\def\bZ{{\bf Z}}
\def\bR{{\bf R}}

\def\bX{{\bf X}}
\def\bY{{\bf Y}}
\def\bZ{{\bf Z}}

\def\mbR{{\mathbb R}}

\DeclareMathOperator{\prox}{Prox}

\def\etal{\emph{et al. }}
\def\ie{\emph{i.e. }}

\usepackage[pagebackref=true,breaklinks=true,letterpaper=true,colorlinks,bookmarks=false]{hyperref}

\cvprfinalcopy 

\def\cvprPaperID{68} 

\ifcvprfinal\fi
\begin{document}

\title{Deep Generative Adversarial Residual Convolutional Networks for Real-World Super-Resolution}

\newcommand{\aand}{\hspace{6mm}}
\author{Rao Muhammad Umer \aand Gian Luca Foresti \aand Christian Micheloni\\
		University of Udine, Italy. \\
		{\tt\small engr.raoumer943@gmail.com, \{gianluca.foresti, christian.micheloni\}@uniud.it}
	}

\maketitle
\thispagestyle{empty}

\begin{abstract}
 Most current deep learning based single image super-resolution (SISR) methods focus on  designing deeper / wider models to learn the non-linear mapping between low-resolution (LR) inputs and the high-resolution (HR) outputs from a large number of paired (LR/HR) training data. They usually take as assumption that the LR image is a bicubic down-sampled version of the HR image. However, such degradation process is not available in real-world settings \ie inherent sensor noise, stochastic noise, compression artifacts, possible mismatch between image degradation process and camera device. It reduces significantly the performance of current SISR methods due to real-world image corruptions. To address these problems, we propose a deep Super-Resolution Residual Convolutional Generative Adversarial Network (SRResCGAN\footnote{Our code and trained models are publicly available at \url{https://github.com/RaoUmer/SRResCGAN}}) to follow the real-world degradation settings by adversarial training the model with pixel-wise supervision in the HR domain from its generated LR counterpart. The proposed network exploits the residual learning by minimizing the energy-based objective function with powerful image regularization and convex optimization techniques. We demonstrate our proposed approach in quantitative and qualitative experiments that generalize robustly to real input and it is easy to deploy for other down-scaling operators and mobile/embedded devices.     
\end{abstract}

\section{Introduction}
The goal of the single image super-resolution (SISR) is to recover the high-resolution (HR) image from its low-resolution (LR) counterpart. SISR problem is a fundamental low-level vision and image processing problem with various practical applications in  satellite imaging, medical imaging, astronomy, microscopy imaging, seismology, remote sensing, surveillance, biometric, image compression, etc. In the last decade, most of the photos are taken using built-in smartphones cameras, where resulting LR image is inevitable and undesirable due to their physical limitations. It is of great interest to restore sharp HR images because some captured moments are difficult to reproduce. On the other hand, we are also interested to design low cost (limited memory and CPU power) camera devices, where the deployment of our deep network would be possible in practice. Both are the ultimate goals to the end users. 
\begin{figure}[!t]
    \centering
    \includegraphics[width=8.0cm]{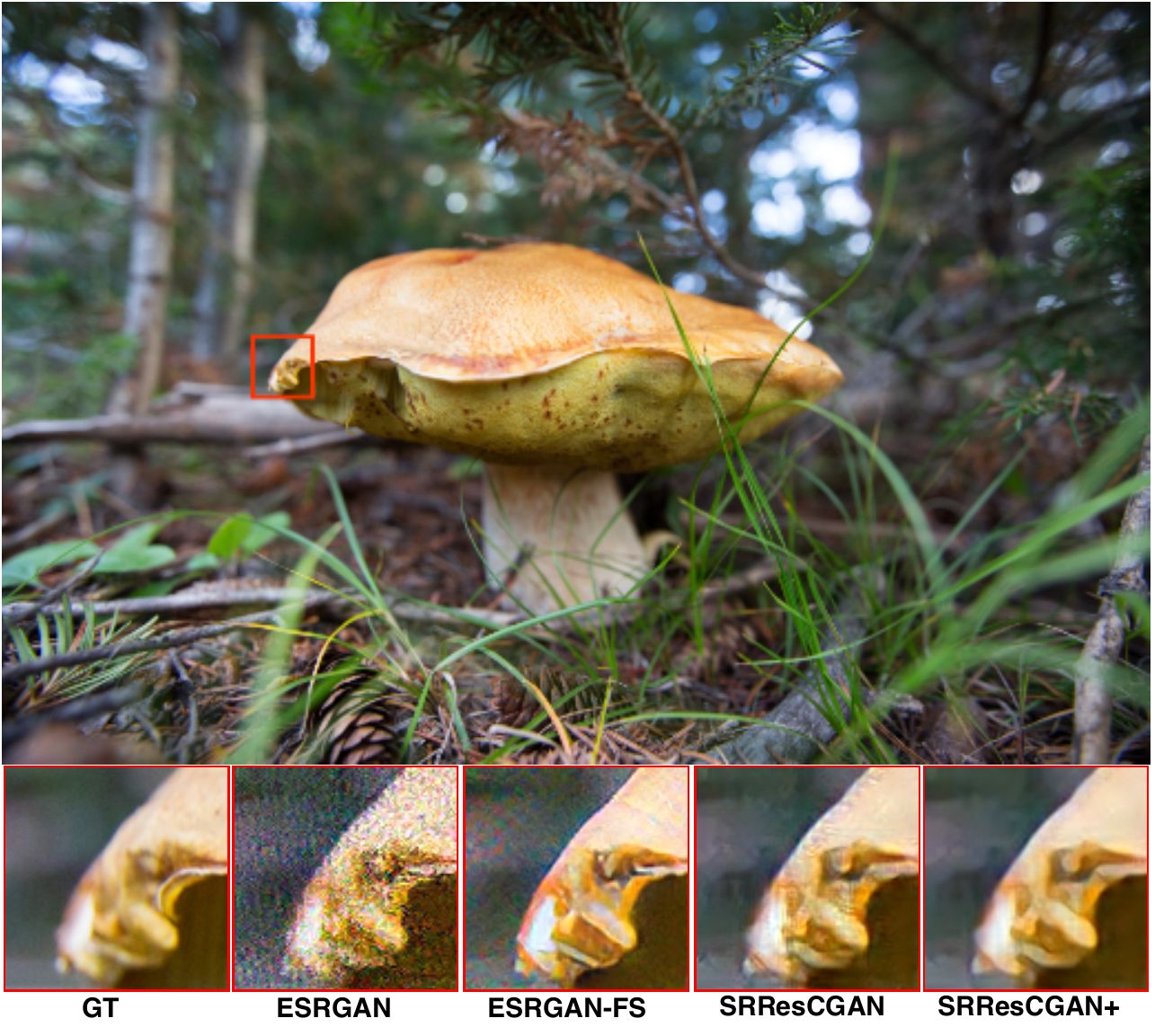}
    \caption{$\times4$ Super-resolution comparison of the proposed SRResCGAN method with the ESRGAN~\cite{wang2018esrgan} and ESRGAN-FS~\cite{fritsche2019dsgan} by the unknown artifacts for the `0815' image (DIV2K validation-set). Our method has better results to handle sensor noise and other artifacts, while the others have failed to remove these artifacts.}
    \label{fig:intro_pic}    
    \vspace{-0.5cm}
\end{figure}

\begin{figure}[h!]
\centering
\includegraphics[scale=0.85]{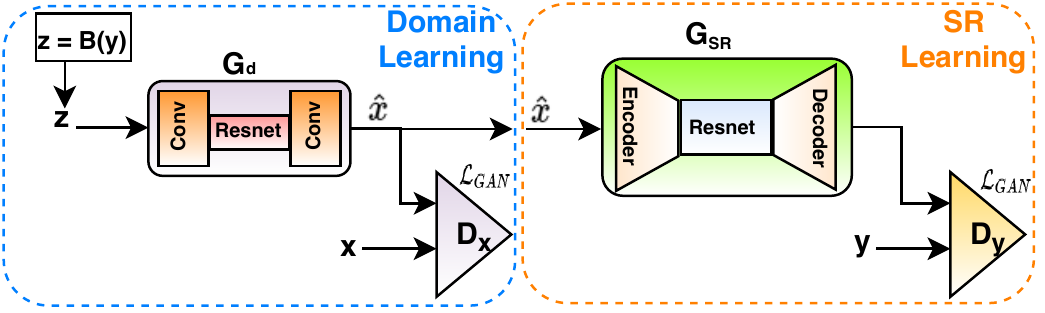}
\caption{Visualizes the structure of the our proposed SR approach setup. In the Domain Learning part, we learn the domain distribution corruptions in the source domain ($\bx$) by the network $\mathbf{G}_d$ (DSGAN structure), where our goal is to map images from $\bz$ to $\bx$, while preserving the image content. Here $\mathbf{B}$ denotes the bicubic downscaling operator which is applied on the clean HR target domain ($\by$) images. In the SR Learning part, we trained the network $\mathbf{G}_{SR}$ in a GAN framework by using generated LR ($\hat{\bx}$) images from the $\mathbf{G}_d$ network with their corresponding HR images.}
\label{fig:srrescgan}
\vspace{-0.5cm}
\end{figure}
Usually, SISR is described as a linear forward observation model by the following image degradation process:
\begin{equation}
    \bY = \bH \Tilde{\bX} + \eta,
    \label{eq:degradation_model}
\end{equation}
where $\bY \in \mbR^{N/s \times N/s}$ is an observed LR image (here $N\times N$ is typically the total number of pixels in an image), $\bH \in \mbR^{N/s \times N/s}$ is a \emph{down-sampling operator} (usually bicubic, circulant matrix) that resizes an HR image $\Tilde{\bX} \in \mbR^{N \times N}$ by a scaling factor $s$ and $\eta$ is considered as an additive white Gaussian noise with standard deviation $\sigma$. But in real-world settings,  $\eta$ also accounts for all possible errors during image acquisition process that include inherent sensor noise, stochastic noise, compression artifacts, and the possible mismatch between the forward observation model and the camera device. The operator $\bH$ is usually ill-conditioned or singular due to the presence of unknown noises ($\eta$) that makes the SISR to highly ill-posed nature of inverse problems. Since, due to ill-posed nature, there are many possible solutions, regularization is required in order to select the most plausible ones.

Since there are many visible corruptions in the real-world images~\cite{lugmayr2019unsupervised, NTIRE2020RWSRchallenge}, current state-of-art SISR methods often fail to produce convincing SR results as shown in Figure~\ref{fig:intro_pic}. Most of the existing SR methods rely on the known degradation operators such as bicubic with paired LR and HR images in the supervised training, while other methods do not follow the image observation (physical) model (refers to Eq.~\eqref{eq:degradation_model}). Three major problems arise in existing SR methods: \textbf{(1)} the first is to train the deeper/wider (lots of model's parameters) networks from a huge volume of training data, \textbf{(2)} the second is not to generalize well for natural image characteristics due to follow the known bicubic down-sampling degradation, and \textbf{(3)} it is not easy to deploy to current generations of smartphone cameras due to lots of network parameters and memory footprints. Therefore, we focus on a robust SISR method that is useful to improve the quality of images in such real-world settings.

In this work, we propose SR learning method (SRResCGAN) that strictly follows the image observation (physical) model (refers to Eq.~\eqref{eq:degradation_model}) to overcome the challenges of real-world super-resolution and is inspired by powerful image regularization and large-scale optimization techniques to solve general inverse problems (\ie easy to deployable for other down-scaling operators). The visualization of our proposed SISR approach setup is shown in Figure \ref{fig:srrescgan}. Due to the unavailability of paired (LR/HR) data, we train firstly the down-sampling (DSGAN)~\cite{fritsche2019dsgan} network ($\mathbf{G}_d$) to generate LR images with same characteristics as the corrupted source domain ($\bx$). We aim to learn the distribution (real-world) mapping from bicubically down-sampled images ($\bz$) of HR images ($\by$) to the source domain images ($\bx$), while preserving the image content. In the second part, the SR network ($\mathbf{G}_{SR}$) is trained in a GAN framework~\cite{goodfellow2014gan} by using generated LR ($\hat{\bx}$) images with their corresponding HR images with pixel-wise supervision in the clean HR target domain ($\by$).   

We evaluate our proposed SR method on multiple datasets with synthetic and natural image corruptions. We use the Real-World Super-resolution (RWSR) dataset~\cite{NTIRE2020RWSRchallenge} to show the effectiveness of our method through quantitative and qualitative experiments. Finally, we also participated in the NTIRE2020 RWSR challenges (track-1 and track-2) associated with the CVPR 2020 workshops. Table~\ref{tab:track1} shows the final testset results of the track-1 of our method (\textbf{MLP-SR}) with others, while we only provide the visual comparison of the track-2 since no ground-truth (GT) is available (refers to Fig.~\ref{fig:4x_result_mobile}), and the quantitative results of the track-2 are in the challenge report~\cite{NTIRE2020RWSRchallenge}. 

\section{Related Work}
\subsection{Image Super-Resolution methods}
Recently, the numerous works have addressed the task of SISR using deep CNNs for their powerful feature representation capabilities. A preliminary CNN-based method to solve SISR is a super-resolution convolutional network with three layers (SRCNN)~\cite{dong2014srcnneccv}. Kim~\etal\cite{kim2016vdsrcvpr} proposed a very deep SR (VDSR) network with residual learning approach. The efficient sub-pixel convolutional network (ESPCNN)~\cite{Shi2016pixelcnncvpr} was proposed to take bicubicly LR input and introduced an efficient sub-pixel convolution layer to upscale the LR feature maps to HR images at the end of the network. Lim \etal\cite{Lim2017edsrcvprw} proposed an enhanced deep SR (EDSR) network by taking advantage of the residual learning. Zhang~\etal\cite{kai2017ircnncvpr} proposed iterative residual convolutional network (IRCNN) to solve SISR problem by using a plug-and-play framework. Zhang~\etal\cite{kai2018srmdcvpr} proposed a deep CNN-based super-resolution with multiple degradation (SRMD). Yaoman \etal\cite{Li2019srfbncvpr} proposed a feedback network (SRFBN) based on feedback connections and recurrent neural network like structure. In \cite{srwdnet}, the authors proposed SRWDNet to solve the joint deblurring and super-resolution task. These methods mostly rely on the PSNR-based metric by optimizing the $\mathcal{L}_1$/$\mathcal{L}_2$ losses with blurry results, while they do not preserve the visual quality with respect to human perception. Moreover, the above mentioned methods are deeper or wider CNN networks to learn non-linear mapping from LR to HR with a large number of training samples, while neglecting the real-world settings.

\subsection{Real-World Image Super-Resolution methods}
For the perception SR task, a preliminary attempt was made by Ledig \etal\cite{wang2018esrgan} who proposed the SRGAN method to produce perceptually more pleasant results. To further enhance the performance of the SRGAN, Wang \etal\cite{wang2018esrgan} proposed the ESRGAN model to achieve the state-of-art perceptual performance. Despite their success, the previously mentioned methods are trained with HR/LR image pairs on the bicubic down-sampling and thus limited performance int real-world settings. More recently, Lugmayr \etal\cite{lugmayr2019unsupervised} proposed a benchmark protocol for the real-wold image corruptions and introduced the real-world challenge series~\cite{AIM2019RWSRchallenge} that described the effects of bicubic downsampling and separate degradation learning for super-resolution. Later on, Fritsche \etal\cite{fritsche2019dsgan} proposed the DSGAN to learn degradation by training the network in an unsupervised way, and also modified the ESRGAN as ESRGAN-FS to further enhance it performance in the real-world settings. However, the above methods still suffer unpleasant artifacts (see the Figures ~\ref{fig:intro_pic}, \ref{fig:4x_result_val} and \ref{fig:4x_result_test}, and the Table~\ref{tab:comp_sota}). Our approach takes into account the real-world settings by greatly increase its applicability. 

\section{Proposed Method}
\subsection{Problem Formulation}
By referencing to the equation~\eqref{eq:degradation_model}, the recovery of $\bX$ from $\bY$ mostly relies on the variational approach for combining the observation and prior knowledge, and is given as the following objective function:
\begin{equation}
    \hat{\mathbf{E}}(\bX) = \underset{\mathbf{X}}{\arg \min } \frac{1}{2}\|\mathbf{Y}-\mathbf{H} \mathbf{X}\|_2^{2}+\lambda \mathbf{R}_W(\mathbf{X}),
    \label{eq:ef1}
\end{equation}
where $\frac{1}{2}\|\mathbf{Y}-\mathbf{H} \mathbf{X}\|_2^2$ is the data fidelity (\ie log-likelihood) term that measures the proximity of the solution to the observations, $\mathbf{R}_W(\mathbf{X})$ is regularization term that is associated with image priors, and $\lambda$ is the trade-off parameter that governs the compromise between data fidelity and regularizer terms.

A generic form of the regularizers in the literature~\cite{chen2017tnrdtpami, Lefkimmiatis2018UDNet} is given as below:
\begin{equation}
    \bR_W(\bX) = \sum_{k=1}^{K} \rho_k(\mathbf{L}_k\bX),
    \label{eq:reg}
\end{equation}
where $\mathbf{L}$ corresponds to the first or higher-order differential linear operators such as gradient, while $\rho(.)$ denotes a potential functions such as $\ell_p$ vector or matrix norms that acts on the filtered outputs \cite{Krishnan2009FastID}. Thanks to the recent advances of deep learning, the regularizer (\ie $\bR_W(\bX)$) is employed by deep convolutional neural networks (ConvNets), whose parameters are denoted by $\mathbf{W}$, that have the powerful image priors capabilities.   

Besides the proper selection of the regularizer and formulation of the objective function, another important aspect of the variational approach is the minimization strategy that will be used to get the required solution. Interestingly, the variational approach has direct link to Bayesian approach and the derived solutions can be described by either as penalized maximum likelihood or as maximum a posteriori (MAP) estimates~\cite{bertero1998map1,figueiredo2007map2}.

\subsection{Objective Function Minimization Strategy}
The proper optimization strategy is employed to find $\mathbf{W}$ that minimizes the energy-based objective function to get the required latent HR image. So, we want to recover the underlying image $\bX$ as the minimizer of the objective function in Eq.~\eqref{eq:ef1} as:  
\begin{equation}
    \hat{\bX} = \underset{\mathbf{X} \in\Tilde{\mathbf{X}}}{\arg \min } ~\hat{\mathbf{E}}(\bX),
    \label{eq:ef2}
\end{equation}
By referencing the Eqs.~\eqref{eq:ef1} and \eqref{eq:reg}, we can write it as:
\begin{equation}
    \hat{\bX} = \underset{\mathbf{X}}{\arg \min } \frac{1}{2}\|\mathbf{Y}-\mathbf{H} \mathbf{X}\|_2^{2}+\lambda \sum_{k=1}^{K} \rho_k(\mathbf{L_k}\mathbf{X}),
    \label{eq:ef3}
\end{equation}
Since it is reasonable to require constraints on the image intensities such as non-negativity values (\ie $\alpha=0,\beta=+\infty$) that arise in the natural images, Eq.~\eqref{eq:ef3} can be rewritten in a constrained optimization form: 
\begin{equation}
    \hat{\bX} = \underset{\alpha \leq \mathbf{X} \leq \beta}{\arg \min } \frac{1}{2}\|\mathbf{Y}-\mathbf{H} \mathbf{X}\|_2^{2}+\lambda \sum_{k=1}^{K} \rho_k(\mathbf{L_k}\mathbf{X}),
    \label{eq:ef4}
\end{equation}
To solve the Eq.~\eqref{eq:ef4}, there are several modern convex-optimization schemes for large-scale machine learning problems such as HQS method~\cite{HQS}, ADMM ~\cite{boyd_admm}, and Proximal methods~\cite{ParikhPGM}. In our work, we solve the under study problem in \eqref{eq:ef4} by using the Proximal Gradient Method (PGM)~\cite{ParikhPGM}, which is a generalization of the gradient descent algorithm. PGM deals with the optimization of a function that is not fully differentiable, but it can be split into a smooth and a non-smooth part. To do so, we rewrite the problem in ~\eqref{eq:ef4} as:
\begin{equation}
    \hat{\bX} = \underset{\mathbf{X}}{\arg \min } \underbrace{\frac{1}{2}\|\mathbf{Y}-\mathbf{H} \mathbf{X}\|_2^{2}+\lambda \sum_{k=1}^{K} \rho_k(\mathbf{L_k}\mathbf{X})}_{\mathbf{F}(\bX)} + \iota_c(\mathbf{X}),
    \label{eq:ef5}
\end{equation}
where $\iota_c$ is the indicator function on the convex set $\mathbb{C} \in \{ \bX \in  \mbR^{N \times N} :\forall k,   \alpha \leqslant \bx_k \leqslant \beta \}$. In \cite{Lefkimmiatis2018UDNet}, Lefkimmiatis proposed a trainable projection layer that computes the proximal map  for the indicator function as:\\
$\iota_c(\mathbf{\bX},~\varepsilon) = \left\lbrace \begin{array}{l}
0~~~~~~, ~~~~~if ~~~\|\bX\|_2 \leqslant \varepsilon \\
+\infty~, ~~~~~otherwise~~~   \\
\end{array}\right.$ \\
where $ \epsilon = e^{\alpha} \sigma \sqrt{C \times H \times W - 1}$ is the parametrized threshold, in which $\alpha$ is a trainable parameter, $\sigma$ is the noise level, and $C \times H \times W$ is the total number of pixels in the image.\\
Thus, the solution of the problem in~\eqref{eq:ef5} is given by the PGM by the following update rule:
\begin{equation}
    \bX^t = \prox_{\gamma^{t}\iota_c}\left(\bX^{(t-1)} - \gamma^t\nabla_{\bX}\mathbf{F}(\bX^{(t-1)})\right),
    \label{eq:ef6}
\end{equation}
where $\gamma^t$ is a step-size and $\prox_{\gamma^{t}\iota_c}$ is the proximal operator~\cite{ParikhPGM}, related to the indicator function $\iota_c$, that is defined as: 
\begin{equation}
     \mathbf{P}_\mathbb{C}(\bZ) = \arg\underset{\bX \in \bC}{\min} ~\frac{1}{2\sigma^2}\|\bX - \bZ\|_2^2 + \iota_c(\mathbf{\bX}),
    \label{eq:ef7}
\end{equation}
The gradient of the $\mathbf{F}(\bX)$ is computed as:
\begin{equation}
    \nabla_{\bX}\mathbf{F}(\bX) = \bH^T(\bH\bX - \bY) + \lambda \sum_{k=1}^{K} \mathbf{L_k}^T\phi_k(\mathbf{L_k}\mathbf{X}),
    \label{eq:ef8}
\end{equation}
where $\phi_k(.)$ is the gradient of the potential function ($\rho_k$). By combining the Eqs.~\eqref{eq:ef6}, ~\eqref{eq:ef7} and \eqref{eq:ef8}, we have the final form of our solution as:
\begin{equation}
    \begin{split}
    \bX^t = \mathbf{P}_\mathbb{C} \left((1-\gamma^t\bH^{T}\bH)\bX^{(t-1)} + ~\gamma^t \bH^T\bY - \right.\\ 
     \left. \lambda \gamma^t \sum_{k=1}^{K} \mathbf{L_k}^T\phi_k(\mathbf{L_k}\mathbf{X^{(t-1)}}) \right),
    \end{split}
    \label{eq:ef9}
\end{equation}
The formulation in~\eqref{eq:ef9} can be thought as performing one proximal gradient descent inference step at starting points $\bY$ and $\bX^{(0)}=0$, which is given by:
\begin{equation}
    \bX = \mathbf{P}_\mathbb{C} \left(\bH^T\bY - 
      \alpha \sum_{k=1}^{K} \mathbf{L_k}^T\phi_k(\mathbf{L_k}\mathbf{Y}) \right),
    \label{eq:ef10}
\end{equation}
where $\alpha = \lambda \gamma$ corresponds to the projection layer trainable parameter, $\mathbf{L_k}^T$ is the adjoint filter of $\mathbf{L_k}$, and $\bH^T$ represents the up-scaling operation.  

\begin{figure*}[h!]
    \centering
    \begin{subfigure}[t]{1.0\textwidth}
        \centering
        \includegraphics[width=12.0cm]{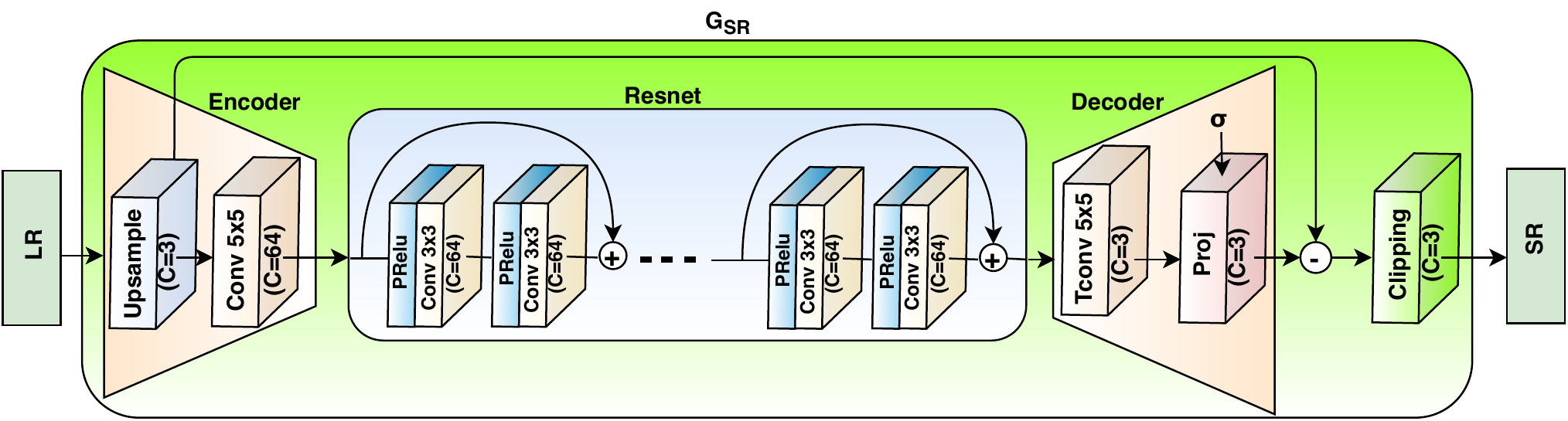}
        \caption{Generator}
    \end{subfigure}\\
    \begin{subfigure}[t]{1.0\textwidth}
        \centering
        \includegraphics[width=12.0cm]{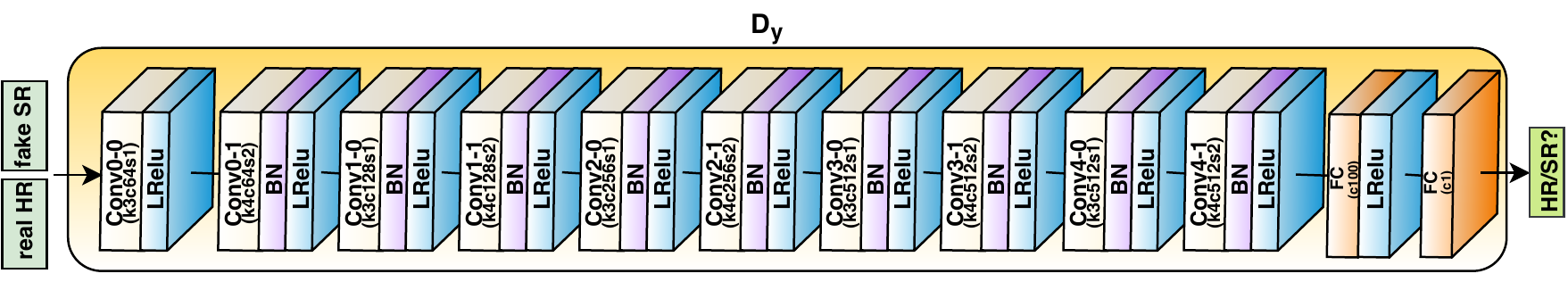}
        \caption{Discriminator.}
    \end{subfigure}
    \caption{The architectures of Generator and Discriminator networks. The $k, c, s$ denotes kernel size, number of filters, and stride size.}
    \label{fig:gen_disc}
\end{figure*}

Thus, we design the generator network ($\mathbf{G}_{SR}$, refers to Fig.~\ref{fig:gen_disc}-(a)) according to Eq.~\eqref{eq:ef10}, where $\phi_k(.)$ corresponds to a point-wise non-linearity (\ie \emph{PReLU}) applied to convolution feature maps. It can be noted that most of the parameters in Eq.~\eqref{eq:ef10} are derived from the prior term of Eq.~\eqref{eq:ef1}, which leads to the proposed generator network as representing the most of its parameters as image priors. In order to learn valid weights of regularization parameters, the weights should be zero-mean and fixed-scale constraints. To tackle this, we use the same parametrization technique proposed in \cite{Lefkimmiatis2018UDNet}. Our generator network structure can also be described as the generalization of one stage TNRD~\cite{chen2017tnrdtpami} and UDNet~\cite{Lefkimmiatis2018UDNet} that have good reconstruction performance for image denoising problem.

\subsection{Domain Learning}
\label{sec:domain_learning}
To learn the domain distribution corruptions from the source domain ($\bx$), we train the network $\mathbf{G}_d$ (see in the Fig.~\ref{fig:srrescgan}) in a GAN framework~\cite{goodfellow2014gan} as done in DSGAN~\cite{fritsche2019dsgan} with the following loss function:
\begin{equation}
    \mathcal{L}_{\mathbf{G}_d} = \mathcal{L}_{color} + 0.005 \cdot \mathcal{L}_{tex} + 0.01 \cdot \mathcal{L}_{per}
    \label{eq:dsgan_lg}
\end{equation}
where, $\mathcal{L}_{color}, \mathcal{L}_{tex}, \mathcal{L}_{per}$ denote the color loss (\ie $\mathcal{L}_1$ loss focus on the low frequencies of the image), texture/GAN loss (\ie focus on the high frequencies on the image), and perceptual (VGG-based) loss, respectively.\\ 
\textbf{Network architectures:} 
The generator network ($\mathbf{G}_d$) consists of 8 \emph{Resnet} blocks (two \emph{Conv} layers and PReLU activations in between) that are sandwiched between two \emph{Conv} layers. All \emph{Conv} layers have $3\times3$ kernel support with 64 feature maps. Finally, \emph{sigmoid} non-linearity is applied on the output of the $\mathbf{G}_d$ network.  While, the discriminator network ($\mathbf{D}_{\bx}$) consists of a three-layer convolutional network that operates on a patch level~\cite{isola2017image, li2016precomputed}. All \emph{Conv} layers have $5\times5$ kernel support with feature maps from 64 to 256 and also applied Batch Norm and Leaky ReLU (LReLU) activations after each \emph{Conv} layer except the last \emph{Conv} layer that maps 256 to 1 features.\\
\textbf{Training description:}
We train the $\mathbf{G}_d$ network with image patches $512\times512$, which are bicubically downsampled with MATLAB \emph{imresize} function. We randomly crop source domain images ($\bx$) by $128\times128$ as do in ~\cite{fritsche2019dsgan}. We train the network for 300 epochs with a batch size of 16 using Adam optimizer~\cite{Kingma2015AdamAM} with parameters $\beta_1 =0.5$, $\beta_2=0.999$, and $\epsilon=10^{-8}$ without weight decay for both generator and discriminator to minimize the loss in \eqref{eq:dsgan_lg}. The learning rate is initially set to $2.10^{-4}$ for first 150 epochs and then linearly decayed to zero after remaining (\ie 150) epochs as do in ~\cite{fritsche2019dsgan}.

\subsection{Super-Resolution Learning}
\label{sec:sr_learning}
\subsubsection{Network Losses}
To learn the super-resolution for the target domain, we train the proposed ($\mathbf{G}_{SR}$) network in a GAN framework~\cite{goodfellow2014gan} with the following loss functions:
\begin{equation}
    \mathcal{L}_{G_{SR}} = \mathcal{L}_{\mathrm{per}}+ \mathcal{L}_{\mathrm{GAN}} + \mathcal{L}_{tv} + 10\cdot \mathcal{L}_{\mathrm{1}}
    \label{eq:l_g}
\end{equation}
where, these loss functions are defined as follows:\\
\textbf{Perceptual loss ($\mathcal{L}_{\mathrm{per}}$):} It focuses on the perceptual quality of the output image and is defined as:
\begin{equation}
    \mathcal{L}_{\mathrm{per}}=\frac{1}{N} \sum_{i}^{N}\mathcal{L}_{\mathrm{VGG}}=\frac{1}{N} \sum_{i}^{N}\|\phi(\mathbf{G}_{SR}(\hat{\bx}_i))-\phi(\by_i)\|_{1}
\end{equation}
where, $\phi$ is the feature extracted from the pretrained VGG-19 network at the same depth as ESRGAN~\cite{wang2018esrgan}.\\
\textbf{Texture loss ($\mathcal{L}_{\mathrm{GAN}}$):} It focuses on the high frequencies of the output image and is defined as:
\begin{equation}
    \begin{split}
       \mathcal{L}_{\mathrm{GAN}} = \mathcal{L}_{\mathrm{RaGAN}} = &-\mathbb{E}_{\by}\left[\log \left(1-\mathbf{D}_{\by}(\by, \mathbf{G}_{SR}(\hat{\bx}))\right)\right] \\
    &-\mathbb{E}_{\hat{\by}}\left[\log \left(\mathbf{D}_{\by}(\mathbf{G}_{SR}(\hat{\bx}), \by)\right)\right] 
    \end{split}
\end{equation}
where, $\mathbb{E}_{\by}$ and $\mathbb{E}_{\hat{\by}}$ represent the operations of taking average for all real ($\by$) and fake ($\hat{\by}$) data in the mini-batches respectively. We employed the relativistic discriminator used in the ESRGAN~\cite{wang2018esrgan} that provides the relative GAN score of real HR and fake SR image patches and is defined as: 
\begin{equation}
   \mathbf{D}_{\by}(\by, \hat{\by})(C) = \sigma(C(\by)-\mathbb{E}[C(\hat{\by})])
\end{equation}
where, $C$ is the raw discriminator output (see in the Fig.~\ref{fig:gen_disc}-(b)) and $\sigma$ is the sigmoid function.\\
\textbf{Content loss ($\mathcal{L}_{\mathrm{1}}$):} It is defined as:
\begin{equation}
    \mathcal{L}_{1} = \frac{1}{N} \sum_{i}^{N} \|\mathbf{G}_{SR}(\hat{\bx}_i)-\by_i\|_{1}
\end{equation}
where, $N$ is represents the size of mini-batch.\\
\textbf{TV (total-variation) loss ($\mathcal{L}_{tv}$):} It focuses to minimize the gradient discrepancy and produce sharpness in the output image and is defined as:
\begin{equation}
    \begin{split}
   \mathcal{L}_{tv}=\frac{1}{N} \sum_{i}^{N}\left(\|\nabla_{h} \mathbf{G}_{SR}\left(\hat{\bx}_{i}\right) - \nabla_{h} \left(\by_{i}\right) \|_{1}+ \right.\\ \left.\|\nabla_{v} \mathbf{G}_{SR}\left(\hat{\bx}_{i}\right) - \nabla_{v} \left(\by_{i}\right) \|_{1}\right)
   \end{split}
\end{equation}
where, $\nabla_{h}$ and $\nabla_{v}$ denote the horizontal and vertical gradients of the images.
\begin{table*}[h!]
	\centering
	\tabcolsep=0.01\linewidth
	\scriptsize
	\resizebox{0.8\textwidth}{!}{
	\begin{tabular}{|l|l|c|c|c|c|c|}
		Dataset (HR/LR pairs) & SR methods & \#Params & PSNR$\uparrow$ & SSIM$\uparrow$ & LPIPS$\downarrow$ & Artifacts \\ 
		\hline
		Bicubic & EDSR~\cite{Lim2017edsrcvprw} & $43M$ & 24.48 & 0.53 & 0.6800 & Sensor noise ($\sigma=8$)\\
		Bicubic & EDSR~\cite{Lim2017edsrcvprw} & $43M$ & 23.75 & 0.62 & 0.5400 & JPEG compression (quality=30)\\
		Bicubic & ESRGAN~\cite{wang2018esrgan} & $16.7M$ & 17.39 & 0.19 & 0.9400 & Sensor noise ($\sigma=8$)\\
		Bicubic & ESRGAN~\cite{wang2018esrgan} & $16.7M$ & 22.43 & 0.58 & 0.5300 & JPEG compression (quality=30)\\
		CycleGAN~\cite{lugmayr2019unsupervised} & ESRGAN-FT~\cite{lugmayr2019unsupervised} & $16.7M$ & 22.42 & 0.55 & 0.3645 & Sensor noise ($\sigma=8$)\\
		CycleGAN~\cite{lugmayr2019unsupervised} & ESRGAN-FT~\cite{lugmayr2019unsupervised} & $16.7M$ & 22.80 & 0.57 & 0.3729 & JPEG compression (quality=30)\\
		DSGAN~\cite{fritsche2019dsgan} &  ESRGAN-FS~\cite{fritsche2019dsgan} & $16.7M$ & 22.52 & 0.52 & 0.3300 & Sensor noise ($\sigma=8$)\\
		DSGAN~\cite{fritsche2019dsgan} & ESRGAN-FS~\cite{fritsche2019dsgan} & $16.7M$ & 20.39 & 0.50 & 0.4200 & JPEG compression (quality=30)\\
		DSGAN~\cite{fritsche2019dsgan} & SRResCGAN (ours) & $380K$ & 25.46 & 0.67 & 0.3604 & Sensor noise ($\sigma=8$)\\
		DSGAN~\cite{fritsche2019dsgan} & SRResCGAN (ours) & $380K$ & 23.34 & 0.59 & 0.4431 & JPEG compression (quality=30)\\
		DSGAN~\cite{fritsche2019dsgan} & SRResCGAN+ (ours) & $380K$ & 26.01 & 0.71 & 0.3871 & Sensor noise ($\sigma=8$)\\
		DSGAN~\cite{fritsche2019dsgan} & SRResCGAN+ (ours) & $380K$ & 23.69 & 0.62 & 0.4663 & JPEG compression (quality=30)\\
		\hline
		DSGAN~\cite{fritsche2019dsgan} & SRResCGAN (ours) & $380K$ & 25.05 & 0.67 & 0.3357 & unknown (validset)~\cite{NTIRE2020RWSRchallenge}\\
		DSGAN~\cite{fritsche2019dsgan} & SRResCGAN+ (ours) & $380K$ & 25.96 & 0.71 & 0.3401 & unknown (validset)~\cite{NTIRE2020RWSRchallenge}\\
		\hline
		DSGAN~\cite{fritsche2019dsgan} & ESRGAN-FS~\cite{fritsche2019dsgan} & $16.7M$ & 20.72 & 0.52 & 0.4000 & unknown (testset)~\cite{AIM2019RWSRchallenge}\\
		DSGAN~\cite{fritsche2019dsgan} & SRResCGAN (ours) & $380K$ & 24.87 & 0.68 & 0.3250 & unknown (testset)~\cite{NTIRE2020RWSRchallenge}\\
	\end{tabular}}
    \caption{Top section: $\times4$ SR quantitative results comparison of our method over the DIV2K validation-set (100 images) with added two known degradation~\ie sensor noise ($\sigma=8$) and JPEG compression ($quality=30$) artifacts. Middle section: $\times4$ SR results with the unknown corruptions in the RWSR challenge track-1 (validation-set)~\cite{NTIRE2020RWSRchallenge}. Bottom section: $\times4$ SR comparison with the unknown corruptions in the RWSR challenge series~\cite{AIM2019RWSRchallenge, NTIRE2020RWSRchallenge}. The arrows indicate if high $\uparrow$ or low $\downarrow$ values are desired.}
	\label{tab:comp_sota}
	\vspace{-0.3cm}
\end{table*}

\subsubsection{Network Architectures}
Figure~\ref{fig:gen_disc} shows the network architectures of both Generator  ($\mathbf{G}_{SR}$) and Discriminator ($\mathbf{D}_{\by}$).\\
\textbf{Generator  ($\mathbf{G}_{SR}$):}
We design the generator network according to the optimization update formula in \eqref{eq:ef10}. In the $\mathbf{G}_{SR}$ network (refers to Fig.~\ref{fig:gen_disc}-(a)), both \emph{Encoder (Conv, refers to $\mathbf{L_k}$ filters)} and \emph{Decoder (TConv, refers to $\mathbf{L_k}^T$ filters)} layers have $64$ feature maps of $5\times5$ kernel size with $C \times H\times W$ tensors, where $C$ is the number of channels of the input image. Inside the \emph{Encoder}, LR image ($\bY$) is upsampled by the Bilinear kernel with \emph{Upsample} layer (refers to the operation $\bH^T\bY$), where the choice of the upsampling kernel is arbitrary. \emph{Resnet} consists of $5$ residual blocks with two Pre-activation \emph{Conv} layers, each of $64$ feature maps with kernels support $3\times3$. The pre-activations (refers to the learnable non-linearity functions  $\phi_k(.)$) are the parametrized rectified linear unit (PReLU) with $64$ out feature channels support. The trainable projection (\emph{Proj}) layer~\cite{Lefkimmiatis2018UDNet} (refers to the proximal operator~$\mathbf{P}_\mathbb{C}$) inside \emph{Decoder} computes the proximal map with the estimated noise standard deviation $\sigma$ and handles the data fidelity and prior terms. Moreover, the \emph{Proj} layer parameter $\alpha$ is fine-tuned during the training via a back-propagation. The noise realization is estimated in the intermediate \emph{Resnet} that is sandwiched between \emph{Encoder} and \emph{Decoder}. The estimated residual image after \emph{Decoder} is subtracted from the LR input image. Finally, the clipping layer incorporates our prior knowledge about the valid range of image intensities and enforces the pixel values of the reconstructed image to lie in the range $[0, 255]$. Reflection padding is also used before all \emph{Conv} layers to ensure slowly-varying changes at the boundaries of the input images.\\
\textbf{Discriminator ($\mathbf{D}_{\by}$):}
The Figure~\ref{fig:gen_disc}-(b) shows the architecture of discriminator network that is trained to discriminate real HR images from generated fake SR images. The raw discriminator network contains 10 convolutional layers with kernels support $3\times3$ and $4\times4$ of increasing feature maps from $64$ to $512$ followed by Batch Norm (BN) and leaky ReLU as do in SRGAN~\cite{ledig2017photo}. 

\subsubsection{Training description}
At the training time, we set the input LR patch size as $32\times32$. We train the network for 51000 training iterations with a batch size of 16 using Adam optimizer~\cite{Kingma2015AdamAM} with parameters $\beta_1 =0.9$, $\beta_2=0.999$, and $\epsilon=10^{-8}$ without weight decay for both generator and discriminator to minimize the loss in \eqref{eq:l_g}. The learning rate is initially set to $10^{-4}$ and then multiplies by $0.5$ after 5K, 10K, 20K, and 30K iterations. The projection layer parameter $\sigma$ is estimated according to \cite{liu2013single} from the input LR image. We initialize the projection layer parameter $\alpha$ on a log-scale values from $\alpha_{max}=2$ to $\alpha_{min}=1$ and then further fine-tune during the training via a back-propagation.
\begin{figure}[htbp!]
    \centering
    \begin{subfigure}[t]{0.5\textwidth}
        \centering
        \includegraphics[width=8.5cm]{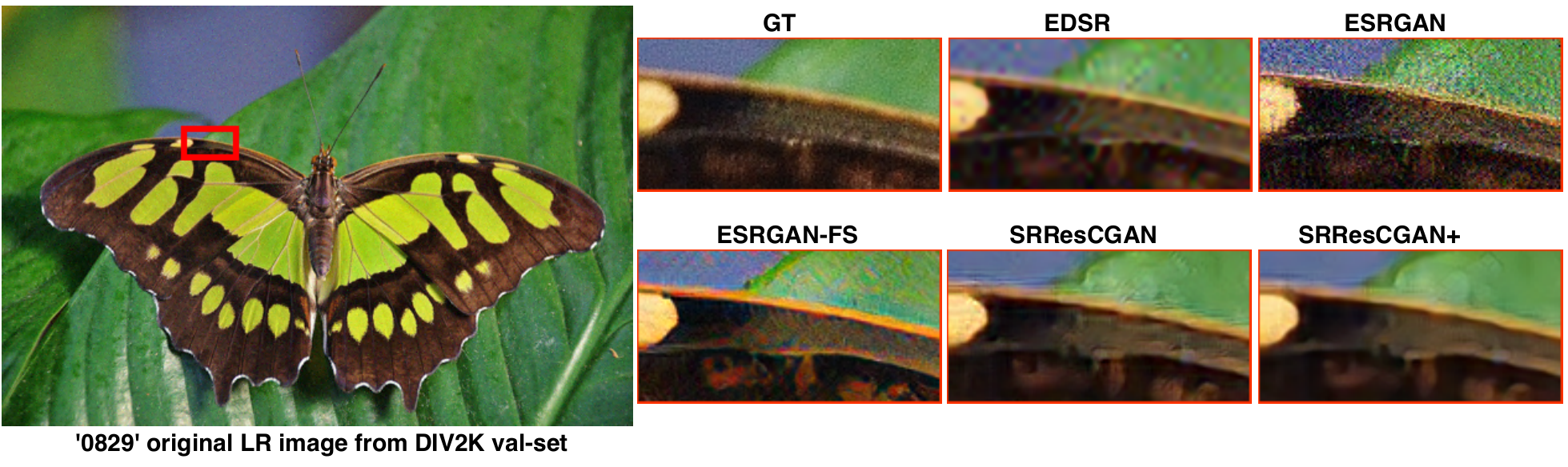}
    \end{subfigure}\\ 
    \begin{subfigure}[t]{0.5\textwidth}
        \centering
        \includegraphics[width=8.5cm]{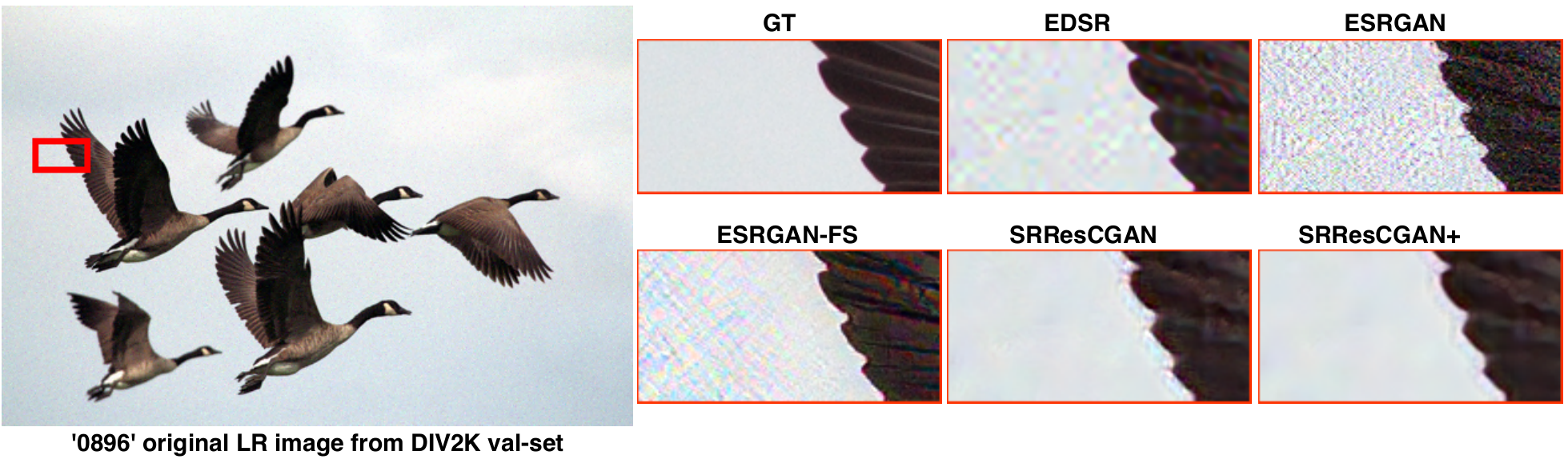}
    \end{subfigure}
    \caption{Visual comparison of our method with other state-of-art methods on the NTIRE2020 RWSR (track-1) validation set~\cite{NTIRE2020RWSRchallenge} at the $\times4$ super-resolution.}
    \label{fig:4x_result_val}
    \vspace{-0.4cm}
\end{figure}
\section{Experiments}
\subsection{Training data}
We use the source domain data ($\bx$: 2650 HR images)  that are corrupted with unknown degradation e.g. sensor noise, compression artifacts, etc. and target domain data ($\by$: 800 clean HR images) provided in the NTIRE2020 Real-World Super-resolution (RWSR)  Challenge track1~\cite{NTIRE2020RWSRchallenge}. We use the source and target domain data for training the $\mathbf{G}_d$ network to learn the domain corruptions, while due to unavailability of paired LR/HR data,  we train the $\mathbf{G}_{SR}$ network (refers to section-\ref{sec:sr_learning}) with generated LR data ($\hat{x}$) from the  $\mathbf{G}_d$ network (refers to section-\ref{sec:domain_learning}) with their corresponding HR target ($\by$) images.

\subsection{Data augmentation}
We take the input LR image patches as generated by the domain learning $\mathbf{G}_d$ network (refers to section~\ref{sec:domain_learning}) with their corresponding HR image patches. We augment the training data with random vertical and horizontal flipping, and $90^{\circ}$ rotations. Moreover, we also consider another effective data augmentation technique, called \emph{MixUp} \cite{zhang2017mixup}. In \emph{Mixup}, we take randomly two samples $(\bx_i , \by_i)$ and $(\bx_j , \by_j)$ in the training LR/HR set $(\Tilde{\bX}, \bY)$ and then form a new sample $(\Tilde{\bx} , \by)$ by interpolations of the pair samples by following the same degradation model \eqref{eq:degradation_model} as do in \cite{feng2019suppressing}. This simple technique encourages our network to support linear behavior among training samples. 

\subsection{Technical details}
We implemented our method with Pytorch. The experiments are performed under Windows 10 with i7-8750H CPU with 16GB RAM and on NVIDIA RTX-2070 GPU with 8GB memory. It takes about 28.57 hours to train the model. The run time per image (on GPU) is  0.1289 seconds at the testset. In order to further enhance the fidelity, we use a self-ensemble strategy~\cite{timofte2016seven} (denoted as SRResCGAN+) at the test time, where the LR inputs are flipped/rotated and the SR results are aligned and averaged for enhanced prediction.

\subsection{Evaluation metrics}
We evaluate the trained model under the Peak Signal-to-Noise Ratio (PSNR), Structural Similarity (SSIM), and LPIPS~\cite{zhang2018unreasonable} metrics. The PSNR and SSIM are distortion-based measures that correlate poorly with actual perceived similarity, while LPIPS better correlates with human perception than the distortion-based/handcrafted measures. As LPIPS is based on the features of pretrained neural networks, so we use it for the quantitative evaluation with features of AlexNet~\cite{zhang2018unreasonable}. The quantitative SR results are evaluated on the $RGB$ color space.

\subsection{Comparison with the state-of-art methods}
\label{sec:comp_sota}
We compare our method with other state-of-art SR methods including EDSR~\cite{Lim2017edsrcvprw}, ESRGAN~\cite{wang2018esrgan}, ESRGAN-FT~\cite{lugmayr2019unsupervised}, and ESRGAN-FS~\cite{fritsche2019dsgan}. Table~\ref{tab:comp_sota} shows the quantitative results comparison of our method over the DIV2K validation-set (100 images) with two known degradation (\ie sensor noise, JPEG compression) as well as unknown degradation in the RWSR challenge series~\cite{AIM2019RWSRchallenge, NTIRE2020RWSRchallenge}. Our method results outperform in term of PSNR and SSIM compared to other methods, while in case of LPIPS, we are slightly behind the ESRGAN-FS ( \ie sensor noise ($\sigma=8$), JPEG compression ($quality=30$)), but ESRGAN-FS has the worst PSNR and SSIM values. We have much better LPIPS ($+0.08$) than the ESRGAN-FS (winner of AIM2019 RWSR challenge~\cite{AIM2019RWSRchallenge}) with unknown artifacts. The ESRGAN-FT has a good LPIPS value, but it achieved the worst PSNR and SSIM scores. Despite that, the parameters of the proposed $\mathbf{G}_{SR}$ network are much less (\ie $\times44$) than the other state-of-art SISR networks, which makes it suitable for deployment in mobile/embedded devices where memory storage and CPU power are limited as well as good image reconstruction quality.

Regarding the visual quality, Fig.~\ref{fig:4x_result_val} shows the qualitative comparison of our method with other SR methods on the $\times 4$ upscaling factor (validation-set). In contract to the existing state-of-art methods, our proposed method produce very good SR results that is reflected in the PSNR/SSIM/LPIPS values, as well as the visual quality of the reconstructed images with almost no visible corruptions. 
\begin{figure}[htbp!]
    \centering
    \begin{subfigure}[t]{0.5\textwidth}
        \centering
        \includegraphics[width=8.5cm]{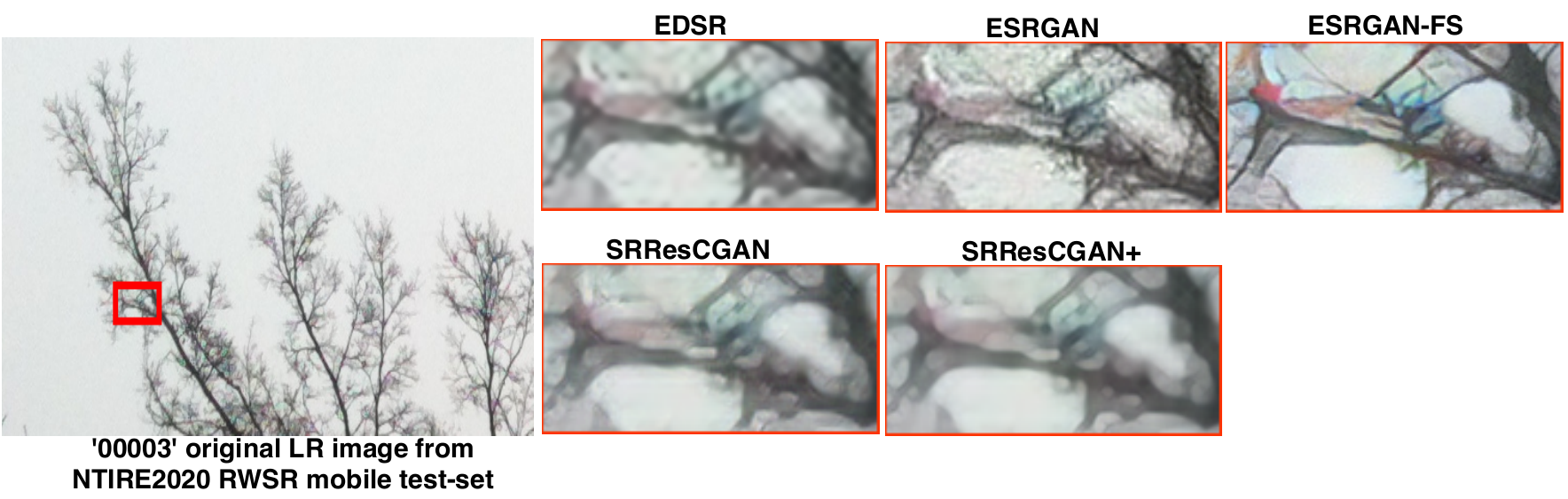}
    \end{subfigure}\\ 
    \begin{subfigure}[t]{0.5\textwidth}
        \centering
        \includegraphics[width=8.5cm]{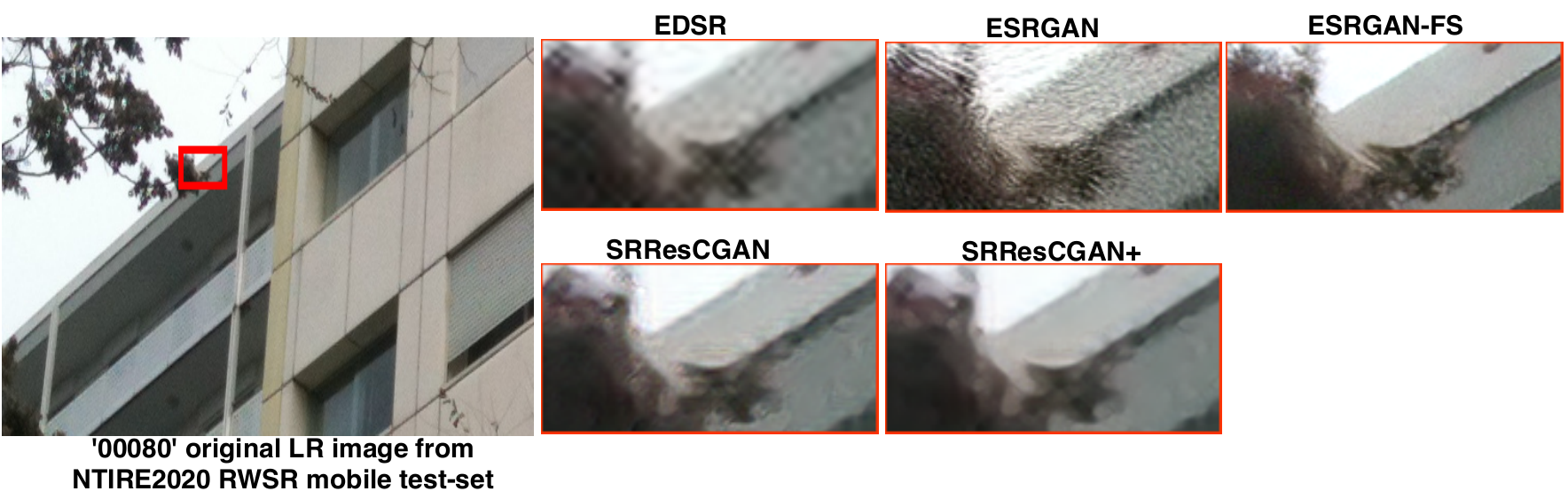}
    \end{subfigure}
    \caption{Visual comparison of our method with other state-of-art methods on the NTIRE2020 RWSR (track-2: Smartphone Images) test set~\cite{NTIRE2020RWSRchallenge} at the $\times4$ super-resolution.}
    \label{fig:4x_result_mobile}
    \vspace{-0.5cm}
\end{figure}

\subsubsection{Visual comparison on the Real-World smartphone images}
We also evaluate our proposed method on the real-world images captured from the smartphone provided in the RWSR challenge track-2~\cite{NTIRE2020RWSRchallenge} (testset). We use the our pretrained model (refers to section-\ref{sec:sr_learning}) without any fine-tuning from the source domain data of the smartphone images for getting the SR results. Since there are no GT images available, we only compare the visual comparison as shown in the Fig.~\ref{fig:4x_result_mobile}. ESRGAN still produces strong noise presence artifacts, while the EDSR produce less noisy, but more blurry results due the PSNR-based metric. ESRGAN-FS produces the sharp images and less amount of corruptions due to fine-tuning of the source domain images (\ie extra training efforts). In contract, our method has still produced satisfying results by reducing the visible corruptions without any extra fine-tuning effort. 
\begin{table}[t]
	\centering%
	\resizebox{0.5\textwidth}{!}{%
		\begin{tabular}{|l|l|l|l|l|}
                     Team &        PSNR$\uparrow$ &        SSIM$\uparrow$ &       LPIPS$\downarrow$ &    MOS$\downarrow$ \\
        \hline
         Impressionism &  24.67 (16) &  0.683 (13) &  0.232 (1) &  2.195 \\
         Samsung-SLSI-MSL &  25.59 (12) &  0.727 (9) &  0.252 (2) &  2.425 \\
         BOE-IOT-AIBD &  26.71 (4) &  0.761 (4) &  0.280 (4) &  2.495 \\
         MSMers &  23.20 (18) &  0.651 (17) &  0.272 (3) &  2.530 \\
         KU-ISPL &  26.23 (6) &  0.747 (7) &  0.327 (8) &  2.695 \\
         InnoPeak-SR &  26.54 (5) &  0.746 (8) &  0.302 (5) &  2.740 \\
         ITS425 &  27.08 (2) &  0.779 (1) &  0.325 (6) &  2.770 \\
         \textbf{MLP-SR} &  24.87 (15) &  0.681 (14) &  0.325 (7) &  2.905 \\
         Webbzhou &  26.10 (9) &  0.764 (3) &  0.341 (9) &  - \\
         SR-DL &  25.67 (11) &  0.718 (10) &  0.364 (10) &  - \\
         TeamAY &  27.09 (1) &  0.773 (2) &  0.369 (11) &  - \\
         BIGFEATURE-CAMERA &  26.18 (7) &  0.750 (6) &  0.372 (12) &  - \\
          BMIPL-UNIST-YH-1 &  26.73 (3) &  0.752 (5) &  0.379 (13) &  - \\
         SVNIT1-A &  21.22 (19) &  0.576 (19) &  0.397 (14) &  - \\
         KU-ISPL2 &  25.27 (14) &  0.680 (15) &  0.460 (15) &  - \\
         SuperT &  25.79 (10) &  0.699 (12) &  0.469 (16) &  - \\
         GDUT-wp &  26.11 (8) &  0.706 (11) &  0.496 (17) &  - \\
         SVNIT1-B &  24.21 (17) &  0.617 (18) &  0.562 (18) &  - \\
         SVNIT2 &  25.39 (13) &  0.674 (16) &  0.615 (19) &  - \\
         \hline
         AITA-Noah-A &  24.65 (-) &  0.699 (-) &  0.222 (-) &  2.245 \\
         AITA-Noah-B &  25.72 (-) &  0.737 (-) &  0.223 (-) &  2.285 \\
         \hline
         Bicubic &  25.48 (-) &  0.680 (-) &  0.612 (-) &  3.050 \\
         ESRGAN Supervised &  24.74 (-) &  0.695 (-) &  0.207 (-) &  2.300 \\
        \end{tabular}
		}
	\caption{Final testset results for the RWSR challenge Track-1. The top section in the table contains ours (\textbf{MLP-SR}) with other methods that are ranked in the challenge. The middle section contains participating approaches that deviated from the challenge rules, whose results are reported for reference but not ranked. The bottom section contains baseline approaches. Participating methods are ranked according to their Mean Opinion Score (MOS).}
	\label{tab:track1}
	\vspace{0mm}
\end{table}

\begin{figure}[htbp!]
    \centering
    \begin{subfigure}[t]{0.5\textwidth}
        \centering
        \includegraphics[width=8.5cm]{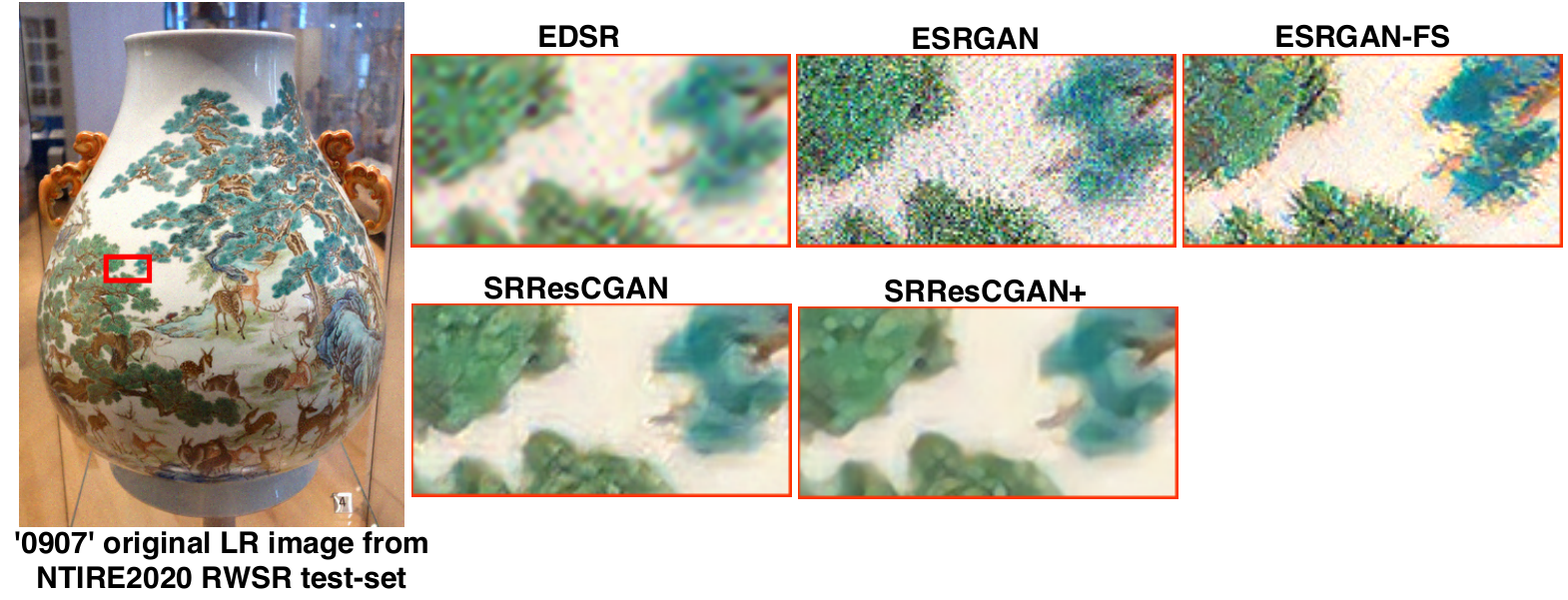}
    \end{subfigure}\\ 
    \begin{subfigure}[t]{0.5\textwidth}
        \centering
        \includegraphics[width=8.5cm]{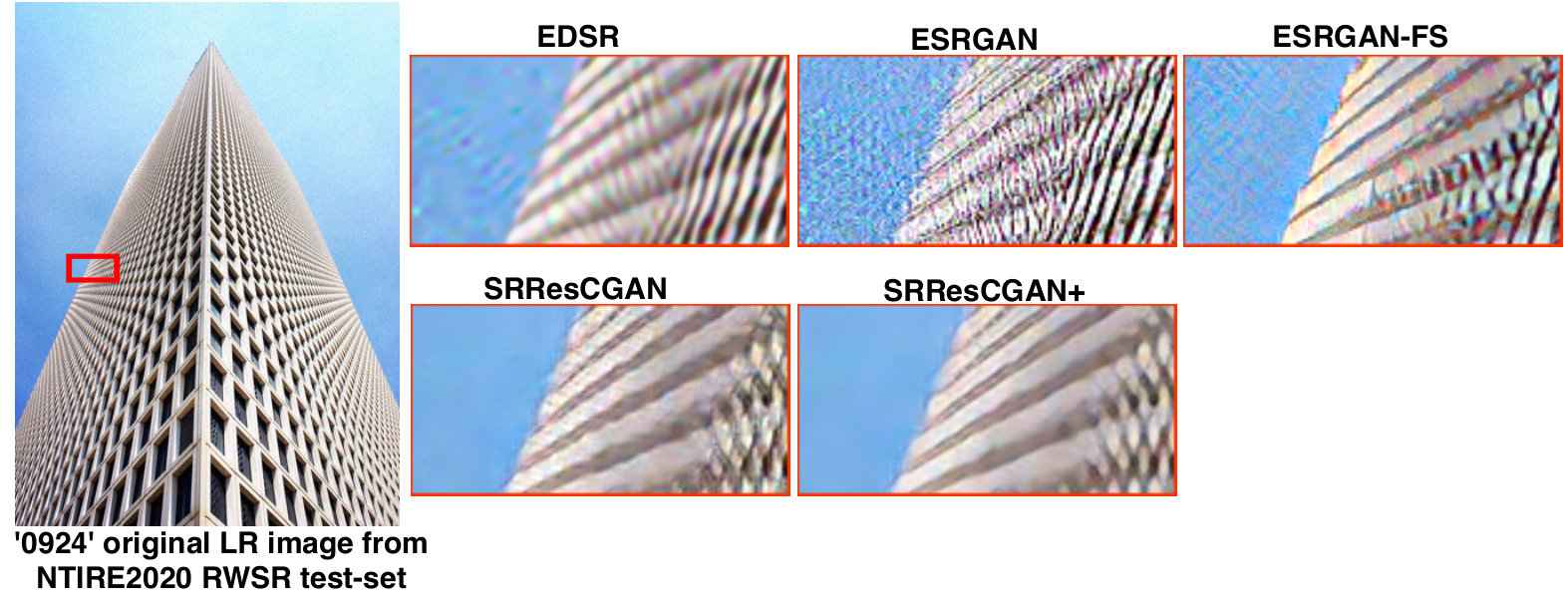}
    \end{subfigure}
    \caption{Visual comparison of our method with other state-of-art methods on the NTIRE2020 RWSR (track-1: Image Processing Artifacts) testset~\cite{NTIRE2020RWSRchallenge} at the $\times4$ super-resolution.}
    \label{fig:4x_result_test}
\end{figure}
\subsubsection{The NTIRE2020 RWSR Challenge}
We also participated in the NTIRE2020 Real-World Super-Resolution (RWSR) Challenge~\cite{NTIRE2020RWSRchallenge} associated with the CVPR 2020 workshops. The goal of this challenge is to super-resolve ($\times4$) images from the Source Domain (corrupted) to the Target Domain (clean). We train firstly the domain learning model ($\mathbf{G}_d$) on the corrupted source domain dataset to learn visible corruptions, and after that train the SR learning model ($\mathbf{G}_{SR}$) on the clean target domain dataset with their correspond generated LR pairs from the ($\mathbf{G}_d$) model (refers to the sections-\ref{sec:domain_learning} and ~\ref{sec:sr_learning} for more details). Table~\ref{tab:track1} provides the final $\times4$ SR results for track-1 (testset) of our method (\textbf{MLP-SR}) with others. The final ranking is based on the Mean Opinion Score (MOS)~\cite{NTIRE2020RWSRchallenge}. Our method remains among the top 8 best solutions. We also provide the visual comparison of our method with others on the track-1 testset in the Fig.~\ref{fig:4x_result_test}. Our method produces sharp images without any visible corruptions, while the others suffer image corruptions. 

\subsection{Ablation Study}
For our ablation study, we compare the different combinations of losses of the proposed SR learning model ($\mathbf{G}_{SR}$). We consider the LPIPS measure for its better visual correlation with the human perception. Table~\ref{tab:ablation_study} shows the quantitative results of our method over the DIV2K validation-set (track-1)~\cite{NTIRE2020RWSRchallenge} with unknown degradation. We first train the SR model with combination of the losses ($\mathcal{L}_{\mathrm{per}}, \mathcal{L}_{\mathrm{GAN}}, \mathcal{L}_{\mathrm{1}}$) similar to ESRGAN. After that, we add $\mathcal{L}_{tv}$ to the previous combinations, and train the model again, we obtain little a bit better LPIPS with sharp SR images. Finally, when we apply the high-pass filter ($\mathbf{w}_{H}$) weights to the output image to compute the GAN loss ($\mathcal{L}_{\mathrm{GAN}}$) focus on the high-frequencies with the previous combinations during training the network, we get the best LPIPS value (\ie $+0.01$ improvement to the previous variants) with more realistic SR images. Therefore, we opt the last one as the final combination of loss functions for our model ($\mathbf{G}_{SR}$) training and also used for evaluation in the section \ref{sec:comp_sota}.  
\begin{table}[!ht]
	\centering
	\tabcolsep=0.01\linewidth
	\scriptsize
	\resizebox{0.50\textwidth}{!}{
	\begin{tabular}{|l|l|c|c|c|}
		\multicolumn{2}{c}{ } & \multicolumn{3}{c}{unknown artifacts} \\
		 SR method & SR Generator Loss combinations ($\mathcal{L}_{G_{SR}}$) & PSNR$\uparrow$ & SSIM$\uparrow$ & LPIPS$\downarrow$ \\ 
		\hline
		SRResCGAN & $\mathcal{L}_{\mathrm{per}}+ \mathcal{L}_{\mathrm{GAN}} + 10\cdot \mathcal{L}_{\mathrm{1}}$ & 25.48 & 0.69 & 0.3458  \\
		SRResCGAN & $\mathcal{L}_{\mathrm{per}}+ \mathcal{L}_{\mathrm{GAN}} + \mathcal{L}_{tv} + 10\cdot \mathcal{L}_{\mathrm{1}}$ & 25.40 &  0.69 & 0.3452  \\
		SRResCGAN & $\mathcal{L}_{\mathrm{per}}+ \mathbf{w}_{H} *\mathcal{L}_{\mathrm{GAN}} + \mathcal{L}_{tv} + 10\cdot \mathcal{L}_{\mathrm{1}}$ & 25.05 & 0.67 & 0.3357  \\
	\end{tabular}}
    \caption{This table reports the quantitative results of our method over the DIV2K validation set (100 images) with unknown degradation for our ablation study. The arrows indicate if high $\uparrow$ or low $\downarrow$ values are desired.}
	\label{tab:ablation_study}
	\vspace{-0.3cm}
\end{table}

\section{Conclusion}
We proposed a deep SRResCGAN method for the real-world super-resolution task by following the image observation (physical / real-world settings) model. The proposed method solves the SR problem in a GAN framework by minimizing the energy-based objective function with the discriminative and residual learning approaches. The proposed method exploits  the powerful image regularization and large-scale optimization techniques for image restoration. Our method achieves very good SR results in terms of the PSNR/SSIM/LPIPS values as well as visual quality compared to the existing state-of-art methods. The proposed method follows the real-world settings for limited memory storage and CPU power requirements (\ie $44$ times less number of parameters than the others) for the mobile/embedded deployment. 
\section*{Acknowledgement}
This work is supported by EU H2020 MSCA through Project
ACHIEVE-ITN (Grant No 765866).
\clearpage
{\small
\bibliographystyle{ieee_fullname}
\bibliography{refs}

\begin{thebibliography}{10}\itemsep=-1pt

\bibitem{bertero1998map1}
Mario Bertero and Patrizia Boccacci.
\newblock {\em Introduction to inverse problems in imaging}.
\newblock CRC press, 1998.

\bibitem{boyd_admm}
Stephen Boyd, Neal Parikh, Eric Chu, Borja Peleato, and Jonathan Eckstein.
\newblock Distributed optimization and statistical learning via the alternating
  direction method of multipliers.
\newblock {\em Foundations and Trends in Machine Learning}, pages 1--122, 2011.

\bibitem{chen2017tnrdtpami}
Yunjin Chen and Thomas Pock.
\newblock Trainable nonlinear reaction diffusion: A flexible framework for fast
  and effective image restoration.
\newblock {\em IEEE Transactions on Pattern Analysis and Machine Intelligence
  (TPAMI)}, pages 1256--1272, 2017.

\bibitem{dong2014srcnneccv}
Chao Dong, Chen~Change Loy, Kaiming He, and Xiaoou Tang.
\newblock Learning a deep convolutional network for image super-resolution.
\newblock {\em ECCV}, pages 184--199, 2014.

\bibitem{feng2019suppressing}
Ruicheng Feng, Jinjin Gu, Yu Qiao, and Chao Dong.
\newblock Suppressing model overfitting for image super-resolution networks.
\newblock {\em IEEE Conference on Computer Vision and Pattern Recognition
  Workshops (CVPRW)}, pages 0--0, 2019.

\bibitem{figueiredo2007map2}
M{\'a}rio Figueiredo, Jos{\'e}~M Bioucas-Dias, and Robert~D Nowak.
\newblock Majorization--minimization algorithms for wavelet-based image
  restoration.
\newblock {\em IEEE Transactions on Image processing}, 16(12):2980--2991, 2007.

\bibitem{fritsche2019dsgan}
Manuel Fritsche, Shuhang Gu, and Radu Timofte.
\newblock Frequency separation for real-world super-resolution.
\newblock {\em ICCV workshops}, 2019.

\bibitem{HQS}
D. Geman and Chengda Yang.
\newblock Nonlinear image recovery with half-quadratic regularization.
\newblock {\em IEEE Transactions on Image Processing (TIP)}, pages 932--946,
  July 1995.

\bibitem{goodfellow2014gan}
Ian Goodfellow, Jean Pouget-Abadie, Mehdi Mirza, Bing Xu, David Warde-Farley,
  Sherjil Ozair, Aaron Courville, and Yoshua Bengio.
\newblock Generative adversarial nets.
\newblock In {\em Advances in neural information processing systems (NIPS)},
  pages 2672--2680, 2014.

\bibitem{isola2017image}
Phillip Isola, Jun-Yan Zhu, Tinghui Zhou, and Alexei~A Efros.
\newblock Image-to-image translation with conditional adversarial networks.
\newblock {\em CVPR}, pages 1125--1134, 2017.

\bibitem{kim2016vdsrcvpr}
Jiwon Kim, Jung~Kwon Lee, and Kyoung~Mu Lee.
\newblock Accurate image super-resolution using very deep convolutional
  networks.
\newblock {\em CVPR}, pages 1646--1654, 2016.

\bibitem{Kingma2015AdamAM}
Diederik~P. Kingma and Jimmy Ba.
\newblock Adam: A method for stochastic optimization.
\newblock {\em ICLR}, pages 1--15, 2015.

\bibitem{Krishnan2009FastID}
Dilip Krishnan and Rob Fergus.
\newblock Fast image deconvolution using hyper-laplacian priors.
\newblock {\em NIPS}, 2009.

\bibitem{ledig2017photo}
Christian Ledig, Lucas Theis, Ferenc Husz{\'a}r, Jose Caballero, Andrew
  Cunningham, Alejandro Acosta, Andrew Aitken, Alykhan Tejani, Johannes Totz,
  Zehan Wang, et~al.
\newblock Photo-realistic single image super-resolution using a generative
  adversarial network.
\newblock {\em CVPR}, pages 4681--4690, 2017.

\bibitem{Lefkimmiatis2018UDNet}
Stamatios Lefkimmiatis.
\newblock Universal denoising networks: A novel cnn architecture for image
  denoising.
\newblock {\em CVPR}, pages 3204--3213, 2018.

\bibitem{li2016precomputed}
Chuan Li and Michael Wand.
\newblock Precomputed real-time texture synthesis with markovian generative
  adversarial networks.
\newblock {\em ECCV}, pages 702--716, 2016.

\bibitem{Li2019srfbncvpr}
Yaoman Li, Jinglei Yang, Zheng Liu, Xiaomin Yang, Gwanggil Jeon, and Wei Wu.
\newblock Feedback network for image super-resolution.
\newblock {\em CVPR}, 2019.

\bibitem{Lim2017edsrcvprw}
Bee Lim, Sanghyun Son, Heewon Kim, Seungjun Nah, and Kyoung~Mu Lee.
\newblock Enhanced deep residual networks for single image super-resolution.
\newblock {\em CVPRW}, pages 1132--1140, 2017.

\bibitem{liu2013single}
Xinhao Liu, Masayuki Tanaka, and Masatoshi Okutomi.
\newblock Single-image noise level estimation for blind denoising.
\newblock {\em IEEE Transactions on Image Processing (TIP)}, pages 5226--5237,
  2013.

\bibitem{lugmayr2019unsupervised}
Andreas Lugmayr, Martin Danelljan, and Radu Timofte.
\newblock Unsupervised learning for real-world super-resolution.
\newblock {\em ICCV workshops}, 2019.

\bibitem{AIM2019RWSRchallenge}
Andreas Lugmayr, Martin Danelljan, Radu Timofte, et~al.
\newblock Aim 2019 challenge on real-world image super-resolution: Methods and
  results.
\newblock In {\em ICCV Workshops}, 2019.

\bibitem{NTIRE2020RWSRchallenge}
Andreas Lugmayr, Martin Danelljan, Radu Timofte, et~al.
\newblock Ntire 2020 challenge on real-world image super-resolution: Methods
  and results.
\newblock {\em CVPR Workshops}, 2020.

\bibitem{ParikhPGM}
Neal Parikh, Stephen Boyd, et~al.
\newblock Proximal algorithms.
\newblock {\em Foundations and Trends in Optimization}, pages 127--239, 2014.

\bibitem{Shi2016pixelcnncvpr}
Wenzhe Shi, Jose Caballero, Ferenc Huszar, Johannes Totz, Andrew~P. Aitken, Rob
  Bishop, Daniel Rueckert, and Zehan Wang.
\newblock Real-time single image and video super-resolution using an efficient
  sub-pixel convolutional neural network.
\newblock {\em IEEE Conference on Computer Vision and Pattern Recognition
  (CVPR)}, pages 1874--1883, 2016.

\bibitem{timofte2016seven}
Radu Timofte, Rasmus Rothe, and Luc Van~Gool.
\newblock Seven ways to improve example-based single image super resolution.
\newblock In {\em CVPR}, pages 1865--1873, 2016.

\bibitem{srwdnet}
Rao~Muhammad Umer, Gian~Luca Foresti, and Christian Micheloni.
\newblock Deep super-resolution network for single image super-resolution with
  realistic degradations.
\newblock In {\em ICDSC}, pages 21:1--21:7, September 2019.

\bibitem{wang2018esrgan}
Xintao Wang, Ke Yu, Shixiang Wu, Jinjin Gu, Yihao Liu, Chao Dong, Yu Qiao, and
  Chen Change~Loy.
\newblock {ESRGAN}: Enhanced super-resolution generative adversarial networks.
\newblock In {\em Proceedings of the European Conference on Computer Vision
  (ECCV)}, 2018.

\bibitem{zhang2017mixup}
Hongyi Zhang, Moustapha Cisse, Yann~N Dauphin, and David Lopez-Paz.
\newblock {Mixup}: Beyond empirical risk minimization.
\newblock {\em International Conference on Learning Representations (ICLR)},
  2018.

\bibitem{kai2017ircnncvpr}
Kai Zhang, Wangmeng Zuo, Shuhang Gu, and Lei Zhang.
\newblock Learning deep cnn denoiser prior for image restoration.
\newblock {\em IEEE Conference on Computer Vision and Pattern Recognition
  (CVPR)}, pages 2808--2817, 2017.

\bibitem{kai2018srmdcvpr}
Kai Zhang, Wangmeng Zuo, and Lei Zhang.
\newblock Learning a single convolutional super-resolution network for multiple
  degradations.
\newblock {\em IEEE Conference on Computer Vision and Pattern Recognition
  (CVPR)}, pages 3262--3271, 2018.

\bibitem{zhang2018unreasonable}
Richard Zhang, Phillip Isola, Alexei~A Efros, Eli Shechtman, and Oliver Wang.
\newblock The unreasonable effectiveness of deep features as a perceptual
  metric.
\newblock {\em IEEE Conference on Computer Vision and Pattern Recognition
  (CVPR)}, pages 586--595, 2018.

\end{thebibliography}
}

\end{document}